\documentclass[aps,prl,reprint,superscriptaddress,showpacs]{revtex4-1}

\usepackage{graphicx}

\begin{document}

\title{Quantized Thermal Conductance of Nanowires at Room Temperature due to Surface Phonon-Polaritons}

\author{Jos\'{e} Ordonez-Miranda}
\affiliation{Laboratoire d'\'{E}nerg\'{e}tique Mol\'{e}culaire et Macroscopique, Combustion, UPR CNRS 288, \'{E}cole Centrale Paris, Grande Voie des Vignes, 92295 Ch\^{a}tenay-Malabry, France.}

\author{Laurent Tranchant}
\affiliation{Laboratoire d'\'{E}nerg\'{e}tique Mol\'{e}culaire et Macroscopique, Combustion, UPR CNRS 288, \'{E}cole Centrale Paris, Grande Voie des Vignes, 92295 Ch\^{a}tenay-Malabry, France.}

\author{Beomjoon Kim}
\affiliation{CIRMM, Institute of Industrial Science, University of Tokyo, Japan.}

\author{Yann Chalopin}
\affiliation{Laboratoire d'\'{E}nerg\'{e}tique Mol\'{e}culaire et Macroscopique, Combustion, UPR CNRS 288, \'{E}cole Centrale Paris, Grande Voie des Vignes, 92295 Ch\^{a}tenay-Malabry, France.}

\author{Thomas Antoni}
\affiliation{Laboratoire d'\'{E}nerg\'{e}tique Mol\'{e}culaire et Macroscopique, Combustion, UPR CNRS 288, \'{E}cole Centrale Paris, Grande Voie des Vignes, 92295 Ch\^{a}tenay-Malabry, France.}
\affiliation{\'{E}cole Centrale Paris, Laboratoire de Photonique Quantique et Mol\'{e}culaire, CNRS (UMR 8537), \'{E}cole Normale Sup\'{e}rieure de Cachan, Grande Voie des Vignes, F-92295 Ch\^{a}tenay-Malabry cedex, France.}

\author{Sebastian Volz}
\affiliation{Laboratoire d'\'{E}nerg\'{e}tique Mol\'{e}culaire et Macroscopique, Combustion, UPR CNRS 288, \'{E}cole Centrale Paris, Grande Voie des Vignes, 92295 Ch\^{a}tenay-Malabry, France.}
\email[]{sebastian.volz@ecp.fr}

\date{\today}

\begin{abstract}
Based on the Landauer formalism, we demonstrate that the thermal conductance due to the propagation of surface phonon-polaritons along a polar nanowire is independent of the material characteristics and is given by $\pi^2 k_B^2 T / 3h$. The giant propagation length of these energy carriers establishes that this quantization holds not only for a temperature much smaller than 1 K, as is the case of electrons and phonons, but also for temperatures comparable to room temperature, which can significantly facilitate its observation and application in the thermal management of nanoscale electronics and photonics.
\end{abstract}


\maketitle

Thermal transport in one-dimensional (1D) systems at low temperature has attracted considerable interest over the past few years, due to the striking quantization of their conductance in integer multiples of a universal quantum. For heat conduction, this quantum due to electrons is equal to that of phonons, and its value is $\pi^2 k_B^2 T / 3h$, where $k_B$  and $h$  are the Boltzmann's and Planck's constants, respectively and $T$ is the temperature \cite{1,2,3,21,22}. Therefore, the minimal amount of heat conducted by electrons and phonons is the same. This result has been theoretically described under the Landauer formalism \cite{4,5,6,7} and validated experimentally \cite{8,9}.

1D energy transport occurs in the ballistic regime \cite{2,9}, that is, when the mean free path of the energy carriers is comparable or larger than the material dimensions. Taking into account that the mean free path of electrons and phonons is of a few nanometers upward, for a wide variety of materials at room temperature \cite{10, 11}, and it increases as temperature decreases, the 1D transport due to these energy carriers is usually achieved in nano-sized materials at very low temperature. For instance, a wire of GaAs with a square cross section of $50^{2} \mathrm{nm}^2$ holds 1D heat conduction for temperatures up to 1 K \cite{2}. By contrast, for heat conduction due to surface phonon-polaritons (SPPs), which are evanescent electromagnetic waves generated by the coupling between photons and phonons at the interface between two different media \cite{12,13}, the propagation length (mean free path) is primarily determined by the dielectric permittivity of these media and not directly by temperature \cite{14}. This indicates that the 1D heat conduction of SPPs is not necessarily restricted to low temperatures, as is the case of electrons and phonons. Therefore, if the SPP thermal conductance is also quantized, this quantization could exist even at room temperature.

The purpose of this letter is to theoretically demonstrate that the thermal conductance due to SPPs propagating along a 1D wire is quantized for any temperature comparable to or smaller than room temperature. Given that the SPP contribution to heat conduction increases as the material size is scaled down \cite{14}, the obtained quantization could be observed in polar nanowires.

\begin{figure}
\includegraphics[width=0.22\textwidth]{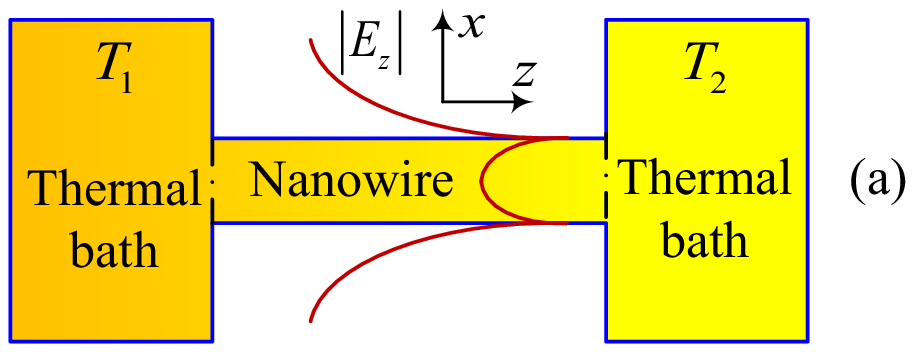}\hfill
\includegraphics[width=0.12\textwidth]{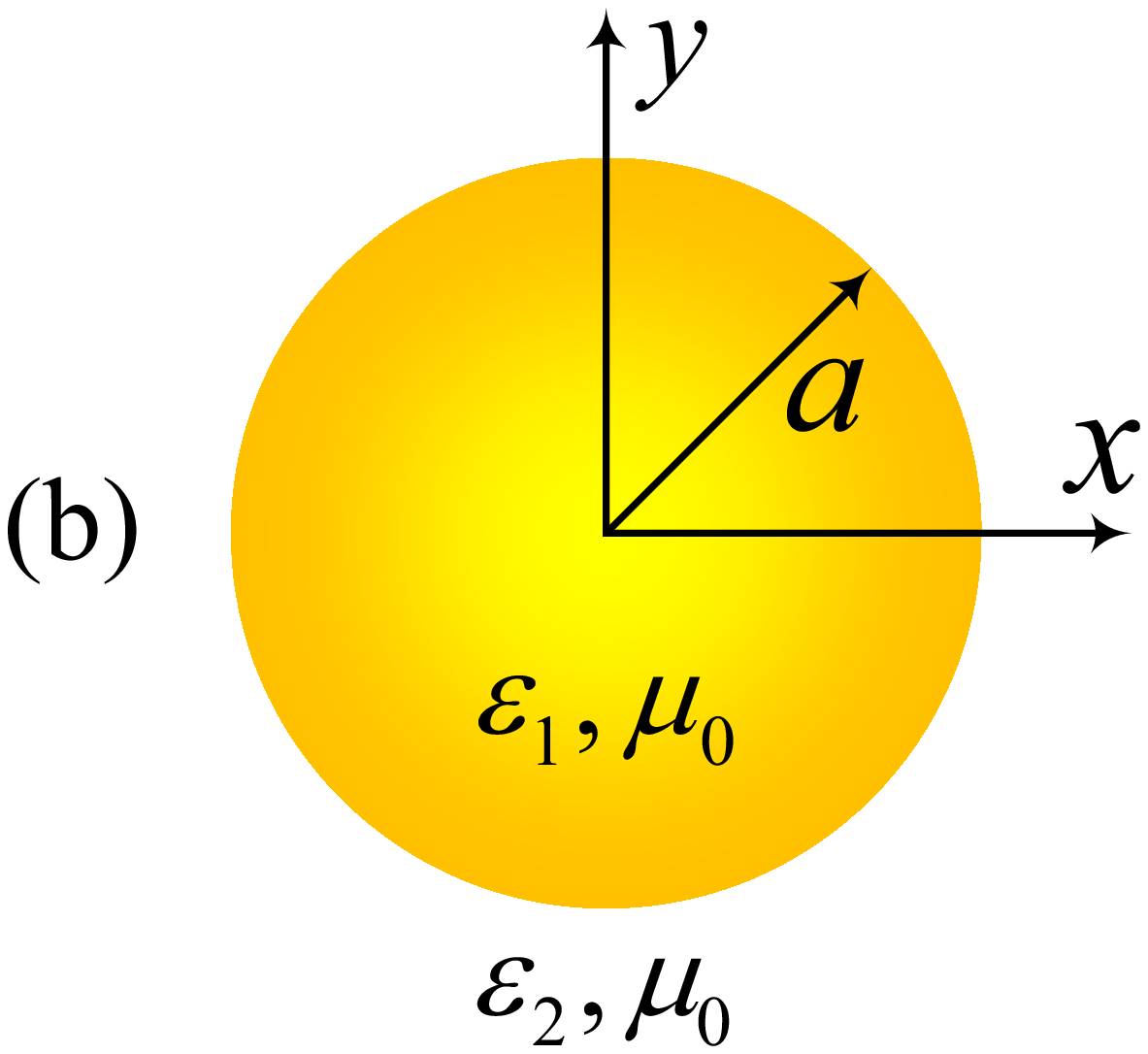}
\caption{\label{fig1}Schematics of (a) the nanowire in thermal contact with two thermal baths and (b) its circular cross-section.}
\end{figure}

Let us consider a wire in thermal contact with two thermal baths set at the temperatures $T_1$  and $T_2$  ( $T_1 > T_2$), as shown in Fig.~\ref{fig1}(a). We first analyze the general 1D heat conduction along this wire and the results are then applied for SPPs. Assuming that the wire is thin enough to conduct heat along its axis mainly (1D wire), the heat flux $Q$  is given by the Landauer formula \cite{5,6}
\begin{equation}
Q=\sum_{n=1}^N \int_0^{k_n^\mathrm{max}} \frac{dk}{2\pi} \hbar \omega V \left[ f_\omega \left(T_1 \right) - f_\omega \left( T_2 \right) \right] \tau \left( k \right),
\label{eq1}
\end{equation}
where $k$ , $\omega = \omega \left(k \right)$  and $V=\partial \omega/\partial k$ are the wave vector, frequency and group velocity of the energy carriers traveling inside the wire; $f_\omega$  is the distribution function of the energy carriers in the thermal baths, $\tau$  is their transmission probability between the nanowire and thermal baths, and  $k_n^\mathrm{max}$ is the largest wave vector at which the modes of the branch $n$ propagate.

Considering that the difference of temperature $\Delta T = T_1 -T_2 << T = (T_1+T_2)/2$, the change of distribution functions in Eq.~(\ref{eq1}) reduces to $f_\omega \left(T_1 \right)-f_\omega \left(T_2 \right)=\Delta T \partial f_\omega \left(T \right) / \partial T$ . The thermal conductance $G=Q / \Delta T$ of the nanowire is then given by
\begin{equation}
G=\frac{1}{2 \pi}\sum_{n=1}^N \int_{\omega_n^\mathrm{min}}^{\omega_n^\mathrm{max}} \hbar \omega \frac{\partial f_\omega\left(T \right)}{\partial T} \tau \left( k \right) d\omega.
\label{eq2}
\end{equation}
The parameters $\omega_n^\mathrm{min}$ and $\omega_n^\mathrm{max}$ stand for the lowest and highest frequencies (cutoff frequencies) of each branch $n$. Equation~(\ref{eq2}) shows that $G$ is independent of all inner details of the dispersion relation $k\left(\omega\right)$, depending only on its cutoff frequencies.  Assuming that the shape effect at the contacts between the nanowire and the thermal baths is small enough to not limit the transmission of energy carriers along the nanowire, we take $\tau = 1$. This can be currently achieved with the high control of growth processes provided by the epitaxial techniques \cite{15}. For SPPs, this assumption is further supported by their very large propagation length \cite{14}, which ensures their transmission through the wire. The SPPs can be thermally generated by heating up the nanowire, through the left thermal bath, to excite its polar molecules, which emit an electric field, as a result of their oscillating electrical dipoles. This field induces the excitation of neighboring electrical dipoles, which keep the propagation of the field (SPP) along the nanowire. To obtain this, the thermal contact between baths and wire should be efficient \cite{14}. For energy carriers following the Bose-Einstein statistics, as is the case of SPPs \cite{16}, Eq.~(\ref{eq2}) can be rewritten as
\begin{equation}
G=\frac{k_B^2T}{h}\sum_{n=1}^N \int_{A_n/T}^{B_n/T} \frac{x^2 e^x}{\left(e^x-1 \right)^2} dx,
\label{eq3}
\end{equation}
where $A_n = \hbar \omega_n^\mathrm{min}/k_B$ and $B_n = \hbar \omega_n^\mathrm{max}/k_B$. The integral in Eq.~(\ref{eq3}) can be calculated analytically and yields
\begin{equation}
G=\frac{k_B^2T}{h}\sum_{n=1}^N \left[J\left(A_n/T \right) - J\left(B_n/T \right)\right],
\label{eq4}
\end{equation}
where
\begin{equation}
J\left(x\right)=\frac{x^2}{e^x-1}-2x \mathrm{ln} \left(1-e^{-x} \right)+2\sum_{m=1}^\infty \frac{e^{-mx}}{m^2}.
\label{eq5}
\end{equation}
Equation~(\ref{eq4}) indicates that the thermal conductance is determined by the relative values of the normalized cutoff frequencies with respect to temperature. The function $J\left( x \right)$ is positive for any $x \ge 0$; for a small argument ($x << 1$), it reduces to  $J\left( x \right) \approx \pi^2 / 3-x$, and for a large argument ($x>>1$), $J\left( x \right) \approx \left[1 + \left(x+1 \right)^2 \right]\exp(-x)$ \cite{23}. To take into account the possible presence of modes with a zero lowest frequency ($\omega_n^\mathrm{min}=0$ ), we define $A_{n}=0$ for the first $N_0$ modes, and $A_{n}>0$ for $n>N_0$. Under this assumption, Eq.~(\ref{eq4}) takes the form
\begin{equation}
G=N_0 G_0+\frac{k_B^2T}{h}\left[\sum_{n=N_0+1}^N J\left(A_n/T\right)-\sum_{n=1}^N J\left(B_n/T \right)\right],
\label{eq6}
\end{equation}
where $G_0=\pi^2 k_B^2 T / 3h$ is the universal quantum of thermal conductance \cite{2}. The following three limiting cases of Eq.~(\ref{eq6}) are of potential interest: 1) for low temperature ($T<<A_n$), the function $J$ in both sums of Eq.~(\ref{eq6}) goes exponentially to zero and hence their contribution is negligible. Thus, Eq.~(\ref{eq6}) reduces to $G=N_0 G_0$, which is the well-known quantization for phonons and electrons at low temperature $T<1 \ \mathrm{K}$ \cite{2,7,9}. 2) For intermediate temperatures ($A_n<<T<<B_n$ ), $J\left(A_n /T \right) \approx \pi^2/3-A_n/T$, $J\left(B_n /T \right) \to 0 $ and Eq.~(\ref{eq6}) reduces to
\begin{equation}
G=N G_0+\frac{k_B}{2\pi}\sum_{n=N_0+1}^N \omega_n^\mathrm{min}.
\label{eq7}
\end{equation}
which depends linearly on the total number of modes $N$ and the temperature through the quantum of thermal conductance. Cases 1) and 2) show that for $T<<B_n$, the thermal conductance exhibits a linear dependence on temperature. 3) For high temperature ($T>>B_n$), Eq.~(\ref{eq6}) yields $G_=G_\infty$, where
\begin{equation}
G_\infty=\frac{k_B}{2\pi}\left(\sum_{n=1}^N \omega_n^\mathrm{max}-\sum_{n=N_0+1}^N \omega_n^\mathrm{min}\right).
\label{eq8}
\end{equation}
Equation~\ref{eq8} establishes that at high temperature, $G$ is independent of $T$ and $G_0$. Its dependence on the material dispersion relation appears only through the difference between the total highest and lowest cutoff frequencies.

\begin{figure}
\includegraphics[width=0.25\textwidth]{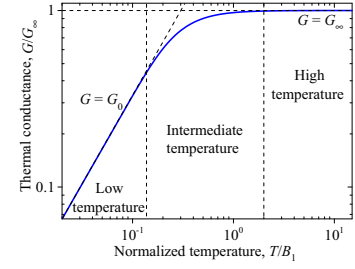}
\caption{\label{fig2}Thermal conductance versus temperature for a nanowire with a single transmitted mode ($N=N_0=1$).}
\end{figure}

Figure~\ref{fig2} shows the thermal conductance $G$ as a function of the normalized temperature for a nanowire with a single transmitted mode. Note that $G$ exhibits three regimes of heat conduction determined by the ratio $T/B_1$, such that its quantization shows up in the low temperature regime ($T<<B_1$). For phonons in a nanowire of GaAs, $B_1 \approx 2 \ \mathrm{K}$ \cite{2}. 

Turning to SPPs propagating along a polar wire with a frequency-dependent permittivity $\varepsilon_1 \left(\omega \right)$, its surrounding medium is considered to have a frequency-independent permittivity $\varepsilon_2>0$, as is the case of air, within a wide range of frequency. Both media are assumed to be non-magnetic ($\mu_{0}=1$). By solving the Maxwell equations under proper boundary conditions for the transverse magnetic polarization required for the existence of SPPs \cite{12,16}, the following dispersion relation for the wave vector $k=\beta$ along the wire axis is obtained \cite{17}
\begin{equation}
\frac{\varepsilon_1}{p_1}\frac{I'_n\left(p_1a\right)}{I_n\left(p_1a\right)}=\frac{\varepsilon_2}{p_2}\frac{K'_n\left(p_2a\right)}{K_n\left(p_2a\right)},
\label{eq9}
\end{equation}
where $I_n$ and $K_n$ are the modified Bessel functions and the prime indicates the derivative. The parameter $n=1,2,...$ accounts for the contribution of the azimuthal modes to the electromagnetic field. The radial wave vectors $p_j$ for the medium $j=1,2$ are given by $p_j^2=\beta^2-\varepsilon_j k_0^2$ where $k_0=\omega/c$; $c$ being the speed of light in vacuum.  Equation~(\ref{eq9}) can significantly be simplified for a thin wire ($ \mid p_j \mid a << 1$), which is of interest in this work to enhance the SPP propagation along the wire. For nanowires of SiC or SiO2, this condition is well satisfied for $a \le 300 \ \mathrm{nm}$ (Fig.~\ref{fig1}(b)). In this case, Eq.~(\ref{eq9}) becomes independent of the radius $a$ and of the azimuthal mode degree $n$, as follows $p_1^2/\varepsilon_1+p_2^2/\varepsilon_2=0$. This indicates that the azimuthal modes does not contribute to the thermal transport through nanowires. The solution of this symmetric relation for $\beta$ is
\begin{equation}
\beta=k_0 \sqrt{2\varepsilon_1\varepsilon_2/\left(\varepsilon_1+\varepsilon_2\right)},
\label{eq10}
\end{equation}
which differs from the dispersion relation of the single plane interface \cite{18}, by just a factor of $\sqrt{2}$, due to the geometry effect. For a nanowire with complex permittivity $\varepsilon_1=\varepsilon_R+i\varepsilon_I$, the real $\beta_R=\xi_{+}$ and imaginary $\beta_I=\xi_{-}$ parts of $\beta$ are given by
\begin{equation}
\xi_{\pm}=\frac{k_0\sqrt{\varepsilon_2}}{\mid\varepsilon_1 + \varepsilon_2 \mid} \sqrt{\mid \varepsilon_1 \mid \mid \varepsilon_1+\varepsilon_2 \mid \pm \mid \varepsilon_1 \mid^2 \pm \varepsilon_R\varepsilon_2},
\label{eq11}
\end{equation}
where it was assumed that $\varepsilon_I>0$. This condition guarantees the absorption of electromagnetic energy by the nanowire, as is the case of real materials. Note that irrespective of the particular complex values of $\varepsilon_1$: 1) the real part $\beta_R>0$, which indicates that the propagation of SPPs is along wire axis, from the hot bath to the cold one. 2) When the frequency tends to zero or infinite, the wire permittivity and hence the square root in Eq.~(\ref{eq10}) tend to a frequency-independent value. Consequently the wave vector $\beta \propto k_0 \to 0$ ($\beta \propto k_0 \to \infty$), that is, $\beta_R\left(\omega=0\right)=0$ ($\beta_R\left(\omega\to\infty\right)\to\infty$). This establishes that the dispersion relation of the SPPs propagating along the wire starts at $A_1=0$ and ends at $B_1\to \infty$ ($N=N_0=1$). 3) Given that $\mid \varepsilon_1\mid\mid\varepsilon_1+\varepsilon_2\mid>\mid\varepsilon_1\mid^2+\varepsilon_R\varepsilon_2$, the imaginary part $\beta_I>0$ and therefore the SPPs propagate with a propagation length $\Lambda=1/2\beta_I$ \cite{14, 16}.

according to Eq.~(\ref{eq10}), the radial wave vectors inside and outside of the nanowire are $p_{1,2}=k_0\sqrt{\varepsilon_{1,2}\left(\varepsilon_{2,1}-\varepsilon_{1,2}\right)/\left(\varepsilon_1+\varepsilon_2\right)}$. The real parts of these wave vectors are positive ($\mathrm{Re}\left(p_j\right)>0$, $j=1,2$), which guarantees that the electrical and magnetic fields decays as they travel away from the nanowire surface. Therefore the SPPs exist for any permittivity $\varepsilon_1$ and frequency $\omega>0$ \cite{14, 20}. Given the existence and propagation of SPPs, Eq.~(\ref{eq6}) and Fig.~\ref{fig2} show that the SPP thermal conductance of a polar nanowire is quantized and equals $G_0$, for any temperature. This value coincides with the one due to electrons and phonons, but for SPPs it not only holds for very low temperatures ($T<1 \ \mathrm{K}$), as is the case of these energy carriers, but also for any higher temperature. The SPP contribution ($G_0$) to the quantum thermal conductance of nanowires at room temperature can then be more than two orders of magnitude higher than the one of phonons, which was measured experimentally with high accuracy, at cryogenic temperatures \cite{9}. This sizeable difference could facilitate the observation of $G_0$.

The obtained quantization is now analyzed for two particular nanowires of SiC and $\mathrm{SiO_2}$ surrounded by air, $\varepsilon_2=1$. For crystalline materials, as SiC, their permittivity is described by the harmonic oscillator model \cite{16}
\begin{equation}
\varepsilon_1(\omega)=\varepsilon_\infty\left(1+\frac{\omega_L^2-\omega_T^2}{\omega_T^2-\omega^2-i\Gamma\omega}\right),
\label{eq13}
\end{equation}
where $\omega_{L}$ and $\omega_{T}$ are the longitudinal and transversal optical frequencies, respectively; $\Gamma$ is a damping constant and $\varepsilon_\infty$ is the high frequency permittivity. The maximum of the imaginary part of $\varepsilon_1$ occurs at $\omega=\omega_T$, for $\omega_T>>\Gamma$, as is the case of most crystals \cite{19}. Therefore, these materials absorb more energy from the electromagnetic field at this frequency. For SiC, $\omega_L=182 \ \mathrm{Trad/s}$, $\omega_T=149 \ \mathrm{Trad/s}$, $\Gamma=0.892 \ \mathrm{Trad/s}$ and $\varepsilon_\infty=6.7$ \cite{19}.

\begin{figure}
\includegraphics[width=0.23\textwidth]{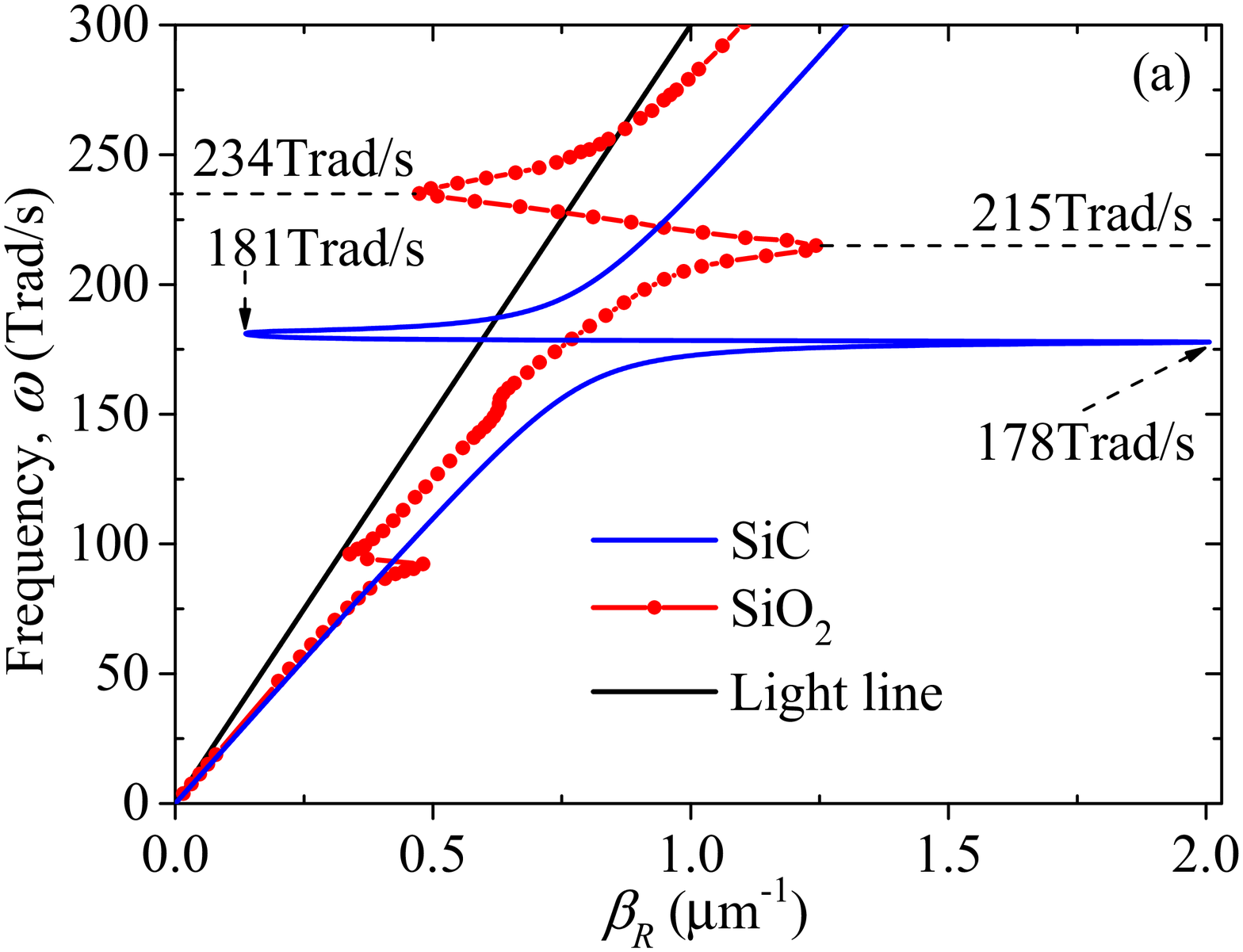}\hfill
\includegraphics[width=0.23\textwidth]{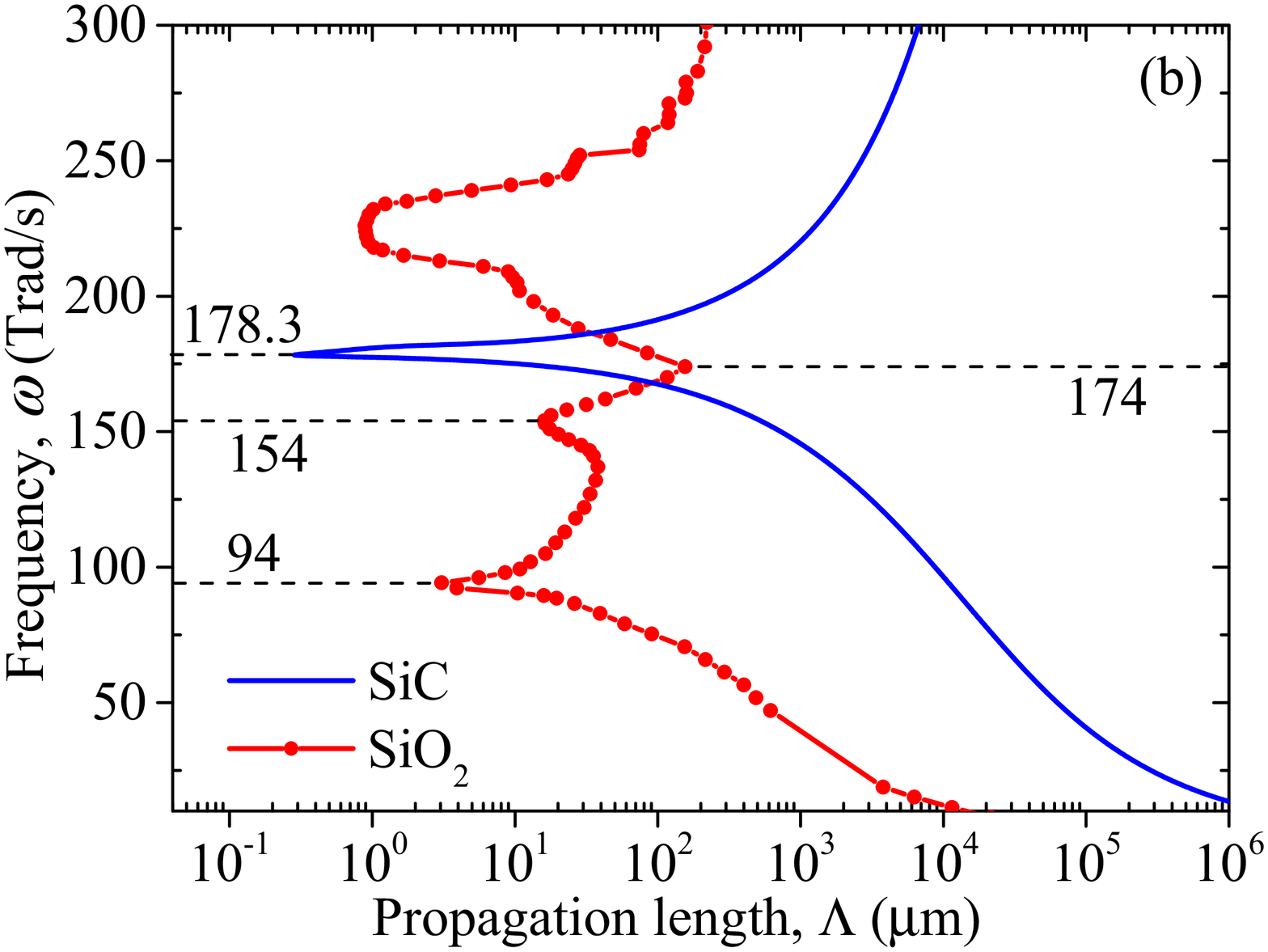}
\caption{\label{fig3}(a) Dispersion relation and (b) propagation length as a function of the frequency, for the nanowires of SiC and $\mathrm{SiO_2}$ surrounded by air ($\varepsilon_2=1$).}
\end{figure}

The dispersion relation and propagation length of SPPs traveling along the nanowire/air interface are shown in Figs.~\ref{fig3}(a) and \ref{fig3}(b), respectively. The curves for the amorphous $\mathrm{SiO_2}$ have been generated using Eq.~(\ref{eq10}) and the experimental data \cite{14} of its complex permittivity. Note that for both SiC and $\mathrm{SiO_2}$, $\beta_R$ decreases and approaches the light line, as the frequency decreases. This indicates a photon-like nature of the SPPs. As the frequency increases, the dispersion curve separates from the light line and tends to a phonon-like behavior. The dispersion curves cross the light line, as a result of the change of sign of 
$\varepsilon_R$, at the frequency $\omega=\omega_L$, for SiC. In contrast to the case of materials with negligible energy absorption ($\varepsilon_I=0$), in the present realistic study ($\varepsilon_I\neq0$), the propagation of SPPs exists here for any wave vector on both sides of the light line, as reported in the literature \cite{20}. In the range of frequencies shown in Fig.~\ref{fig3}(a), the propagation length in Fig.~\ref{fig3}(b) is well defined and takes larger (smaller) values at the frequencies where the absorption of energy is minimum (maximum). In order to ensure the propagation of the SPPs through the entire nanowire, its length has to be smaller or equal to the SPP propagation length, which can be as high as 1cm, as shown in Fig. 3(b).

\begin{figure}
\includegraphics[width=0.23\textwidth]{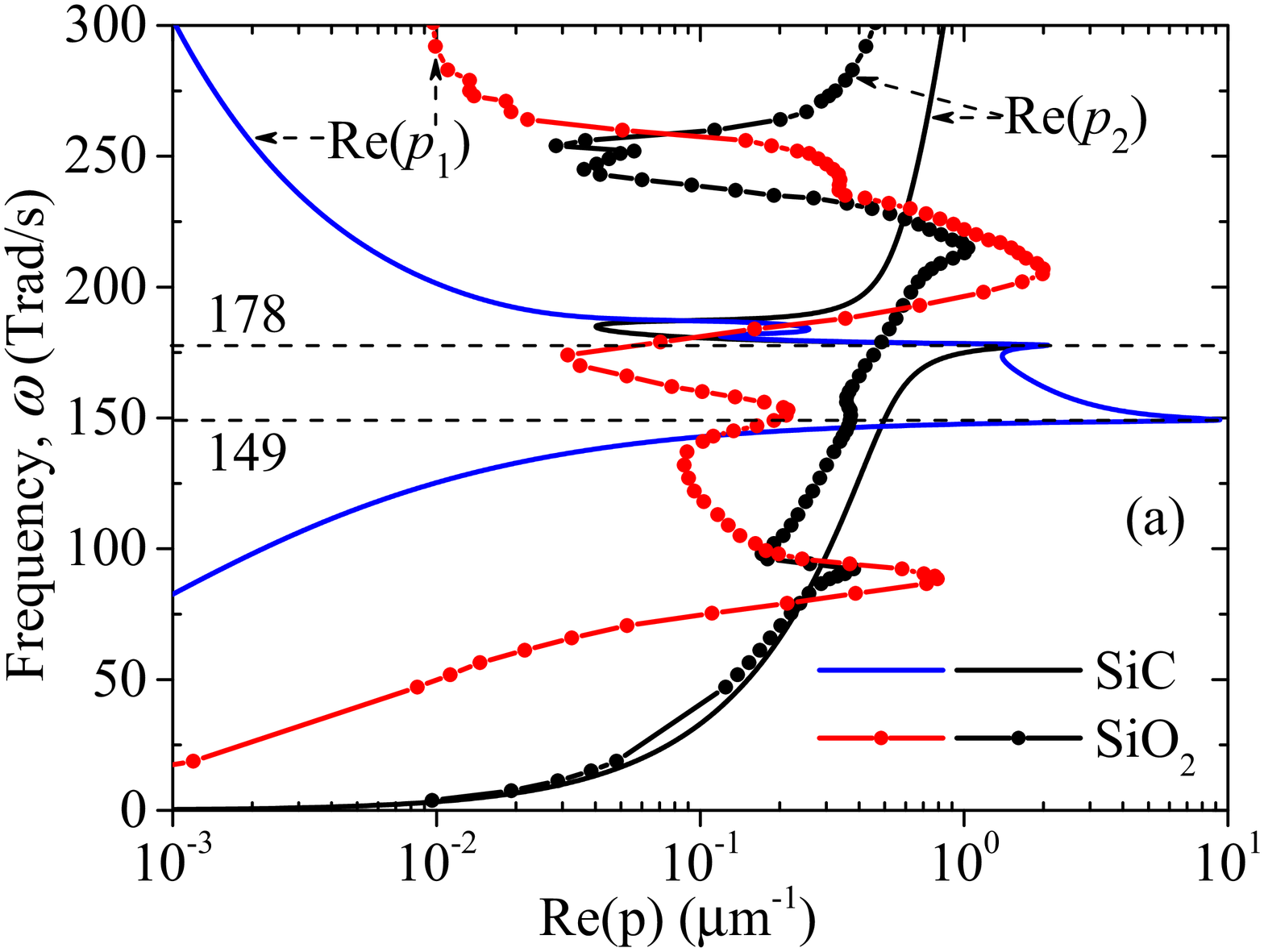}\hfill
\includegraphics[width=0.23\textwidth]{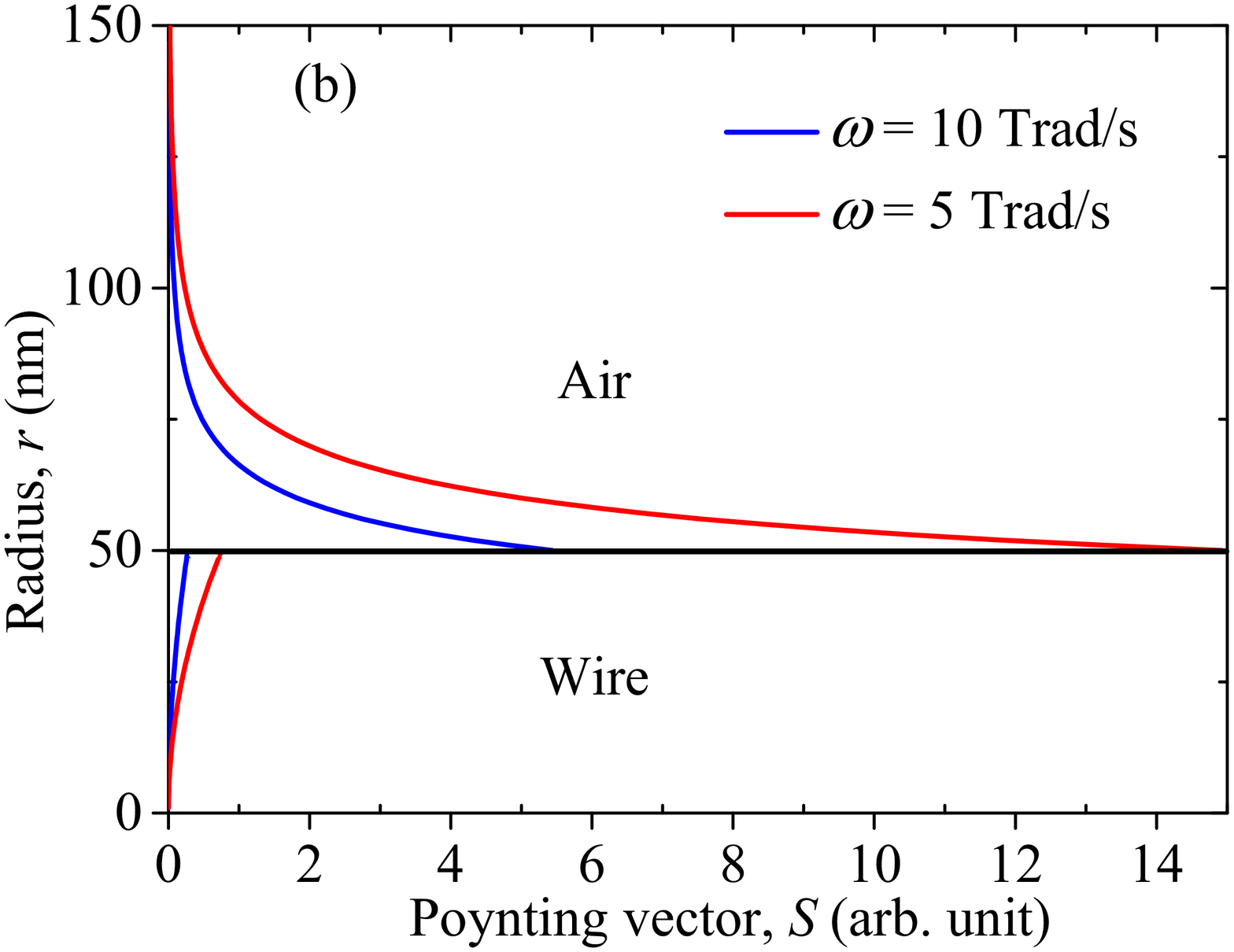}
\caption{\label{fig4}(a) Frequency dependence of Re($p$) inside and outside of the nanowires of SiC and $\mathrm{SiO_2}$ surrounded by air. (b) Poynting vector for a nanowire of SiC with radius $a=50 \ \mathrm{nm}$, as a function of the radial coordinate.}
\end{figure}

The real parts of the radial wave vectors inside and outside of the nanowire are positive, as shown in Fig.~\ref{fig4}(a). This fact along with Figs.~\ref{fig3} reveals that the propagation of SPPs starts at $\omega=0$ and is present in a broad band of frequencies. This is further confirmed by the distribution of the Poynting vector shown on Fig.~\ref{fig4}(b) \cite{17}. Note that the energy flux propagating inside the wire is quite small in comparison with that outside of it, due to the very small absorption by the surrounding air. The high concentration of energy at the interface enables the absorbing nanowire to support the propagation of SPPs. The energy flux increases as the frequency decreases, which indicates that the major contribution to the SPP thermal conductance of the nanowire arises from the low frequency regime.

Given the existence (Figs.~\ref{fig4}) and propagation (Fig.~\ref{fig3}(b)) of SPPs along the nanowire/air interface, Fig.~\ref{fig3}(a) indicates that the dispersion relation for the propagation of SPPs contains one branch ($N=N_1$), which start at the zero frequency ($A_1=0$). Note that $\omega_1^\mathrm{max}>300$ Trad/s, and hence $B_1=\hbar\omega_1^\mathrm{max}/k_B>2283 \ \mathrm{K}$. Thus, Fig.~\ref{fig2} establishes that for any temperature comparable or smaller than 300 K ($T/B_1\to 0$), the SPP thermal conductance of the nanowire of both SiC and $\mathrm{SiO_2}$ is quantized by the value $G_0$. More generally, this quantization holds for any polar nanowire, as rigorously demonstrated above. This is not the case of ideal materials with zero absorption ($\varepsilon_I=0$), in which the existence of SPPs is restricted to narrow intervals of high frequencies, determined by $\varepsilon_1<-\varepsilon_2$ \cite{14,19}. 

The SPP quantum of thermal conductance could be measured generating SPPs by thermal excitation at one side of the nanowire, and detecting their diffraction at the other side. This diffracted signal contains information about the SPP contribution to the heat flux along the nanowire and it can be recorded through an IR microscope, over a wide range of frequencies and temperatures comparable to room temperature. 
Provided that the phonon contribution to the thermal conductance is known, the experimental data are expected to yield a SPP thermal conductance with a linear dependence on the temperature, which is the signature of its quantization (see supplemental material). Given that the quantum of thermal conductance due to phonons is not present at room temperature, the detection of this signature should be due to the SPP contribution only.

In summary, the existence of a universal quantum of thermal conductance $\pi^2 k_B^2 T / 3h$ due to the propagation of SPPs along the surface of a polar nanowire has been demonstrated, not only for temperatures much smaller than 1 K, as is the case of electron and phonons, but also for temperatures comparable to room temperature. This quantization arises from the SPPs propagating with very large propagation lengths at low frequency. The room-temperature result can provide guidelines to improve the thermal performance of nanomaterials.

\bibliography{bibliographie}

\end{document}